# *Quick Safari Through Software Design*

M. Reza Rahimi

## 1. Design Process

By looking at the dictionary [Webster_1998], one could find the following definition of the verb *'to Design*':

> *"to create, execute, or construct according to the plan. According to etymology science it is from the latin word **designare** (**de** + **signare**) which means to mark."*

It could be understood from the above definition that to design something related to *make* artifacts which could be used for special purpose. In brief the purpose of design is to produce a solution to the problem. Another interesting point about the root of the word design is its relation to the word *signature*. One possible way of thinking about it is; the *design* is the *signature* of the *designer*; a person who makes an object. In summary we could define design as the activity that produces any form of artifact [Budgen_2003].

The other words that are usually used in this context are; **engineering**, **engineer** and **engine**. The term **engineering** itself has a much more recent etymology and, deriving from the word **engineer**, which itself dates back to 1325, when an **engine'er**, originally referred to "*a constructor of military engines* "[Oxford_1989]. In this context, an "engine" referred to a military machine, for example, something that is used in war. The word "engine" itself is of even older origin, ultimately deriving from the Latin *ingenium*, which means "innate quality, especially mental power, hence a clever invention. So the engineering could be defined as [Webster_1998]:

> *"The creative application of science and mathematics by which the properties of matter and the sources of energy in nature are made useful to people."*

In any engineering process the basic science should be known ( it is sometime called *domain knowledge* ). One question that may arise is: "Is there any differences or similarities between the process that usually scientists take and the process that engineers or designers take?" To answer this question, we should distinguish between these two processes. The first one is the *scientific process* and the second one is the *engineering process*. Fig. 1 shows the relationship of these two processes.

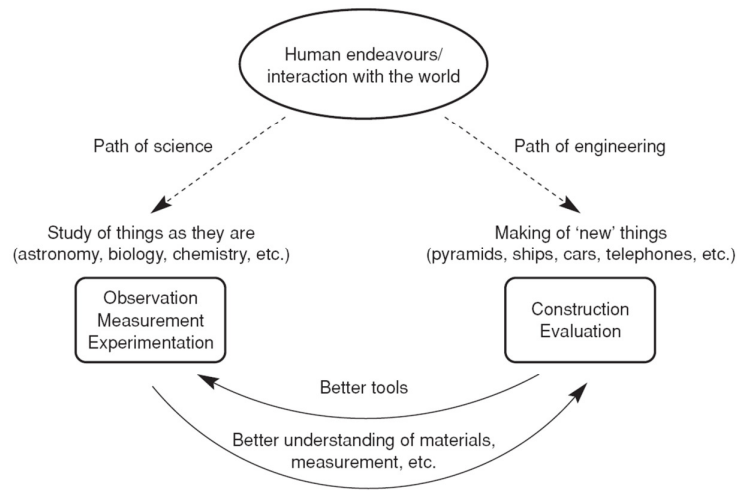

**Fig 1. The Relation between Scientific Process and Engineering Process [Budgen_2003]**

As it can be seen that the human interaction with the world makes two different paths. In the first one which is the path of science, the scientist tries to observe the nature. By measuring some features, he wants to make a model and frame works that could predict the nature and its behavior. It needs more experiences and experiment for proving the validation of the model, and deriving the scientific principles, most probably using mathematical notations.

On the other side is the path of Engineering. By knowing the rules that govern the nature, engineer or designer (In the context of this report I will use the term engineer, engineering and designer, designing as the same.) could clarify the nature of the requirement that he wants to make some artifacts. He uses this requirement and *domain knowledge* ( which could be defined as the specific science and knowledge related to the context, for example when a designer wants to make a bridge he should have knowledge about mechanics, soil,…) to analysis the problem and makes the *black box* model of the problem. The black box is kind of *diagram* or *representation* that describes the external behavior of the element or *what* that element should do. After getting the general *architecture* of the problem, he will change the black box to the *white box* which contains the practical solution of the problem or how the black box should be implemented. The prototyping will be used to verify and validate the design or engineering task. Finally the product will be produced and manufactured. Fig. 2 shows this process.

There are some common points and relations between scientific process and design process:

1. The scientific process and design process are *complementary* processes. It means that the progress in one of this activity could help the progress in the other one. For example engineers could make the exact tools for measuring, and scientist could find the exact rules in nature which help engineers to make a better and more accurate tools.

2. Both of these processes are *iterative* which means that to achieve a *desired result,* repeated cycle of operations should be done. The desired result is depend on the context of the problem domain, for example in science the scientist should verify that if theory could predict the results with accepted tolerance and in design process it could be checked that if the design meets the requirements.
3. Both of these processes use the concept of **abstraction**. Abstraction is the process or result of generalization by reducing the information content of a concept or an observable phenomenon, typically in order to retain only information which is relevant for a particular purpose [Susanne_1953].

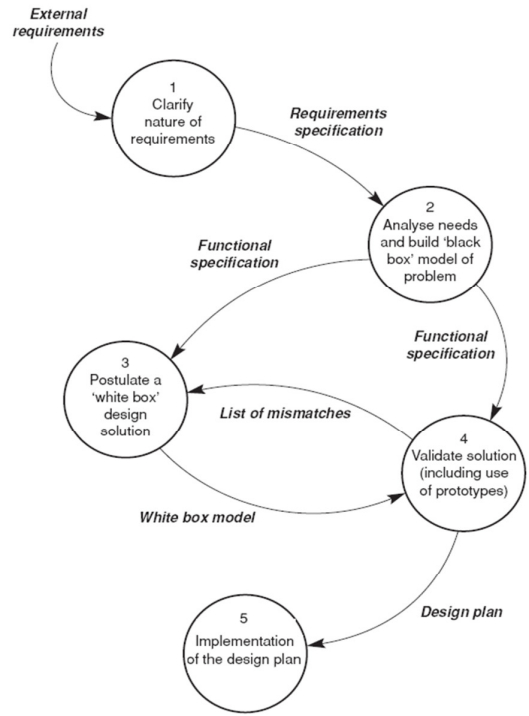

**Fig 2. Model of the Design Process [Budgen_2003]**

These two processes have some different aspects. One of the most important differences between them is that, scientific process usually uses *mathematics* as a powerful *tool* for solving the problem, while the design process is not usually based on mathematical tools (although there exists for some applications). In design process the other *tools* and *techniques,* which **design methods** are usually based on, are used for solving the problem. Some of the common techniques and concepts that are used in almost all design methods are [Budgen_2003]:

1. **Modularity**: Modularity is a general systems concept, typically defined as a continuum describing the degree to which a system's components may be separated and recombined [Schilling_2000]. In general the higher the modularity the better overall quality.
2. **Reuse**: To use the element of the design more than once. If we have elements that are optimized, we could use them easily in the other design. The process of *standardization* is considered as the way to better use of this concept in the engineering task. The *design pattern* could be considered as the design technique in this category.

The design process has the other characteristic that distinguishes it from scientific process. It is very rare that design converges, which means that it directs the designer to the single solution. In this sense the design is sometimes called the *wicked problem* [Budgen_2003]. Wicked problem can be characterized as a problem whose form is such that a solution for one of its aspects simply changes the problem. Wicked problem have some characteristics which could be briefly listed as [Budgen_2003]:

1. There is no definite formulation of a wicked problem.
2. Wicked problem has no stopping rule.
3. Solutions to wicked problems are not true or false, but good or bad.
4. There is no immediate and no ultimate test of a solution to a wicked problem.
5. Every solution to a wicked problem is a 'one-shot operation', because there is no opportunity to learn by trial-and-error, every attempt counts significantly.
6. Wicked problem do not have an enumerable set of potential solutions.
7. Every wicked problem is essentially unique.
8. Every wicked problem can be considered to be a symptom of another problem.

As it is clear from the definition of the wicked problem the design activity is considered to be a challenging task, which needs, creativity, effort, good domain knowledge, etc. In this report I focus on the design process in the context of software engineering. In the next section the software design will be discussed.

# 2. Software Design

In this section I am going to define the software design and try to have an overall classification of the software methods. After that I will derive the ideal method for software design. The software could be defined as [Webster_1998]:

*"The entire set of programs, procedures, and related documentation associated with a system and especially a computer system"*

The software design process could be defined as the process of the problem solving that produces the software. Fig.3 presents the general model and phases in almost every software design process. In the first phase of every design the requirements should be clarified. After this phase the designer design the general or abstract model of the design. This phase is sometimes called *architectural* or *logical* design phase. In this phase the designer only describe the external behavior of each element in the design. The next phase is called the *detailed design* phase. In this phase the architecture or logical block are mapped on to the implementable units, which could be realized using technology. This phase is also called the transition from black boxes to white boxes [Budgen_2003]. It should be noticed that the nature of the design methods is dependent on a number of factors, such as software development environment, software resources, the quality and knowledge of software teams, etc.

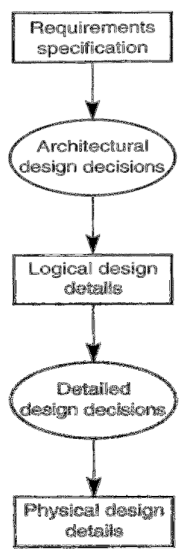

**Fig 3. General phases of Software Design Process [Budgen_2003]**

Although the design methods are sometimes different in nature but they have many common points such as mechanism for translation of information domain representation into design representation, notation for functional component and their interfaces and guidelines for quality assessments [Pressman_2001]. In the next section I am going to classify the design methods, those are proposed for software design or could be used in software design process.

## 2.1 Software Design Methods: Overall Classification

In one way of classification of software methods, one could categorize them into two classes: *Formal Design Methods* and *Systematic or Procedural Design Methods* [Pressman_2001]. The formal methods use extensive knowledge of mathematics and logic in the process of design for the element transformation and verification and validation [Suh_2001]. The systematic types are less mathematical and consist of the procedure components, describes what action or task should be assigned to each component. In the following section I am trying to categories the systematic or procedural methods.

## 2.1.1 Top-Down and Bottom-Up Design Methods

In top-down design, the designers should start from the top-level description of the system and then refine and optimize this view step by step. In each step the designer decompose the higher level components into the lower level until the lower level element implementable. The top-down design approach, reduces the scope and size of each module, and focuses more on specific issues [Yourdon_1979]. The top-down method is an iterative process where each refinement will decompose a module into more specific and detailed sub-modules until it reaches a point where it could be achieved. The decisions made at the upper-levels will have a significant effect on subsequent decomposition at the lower-levels [Thomas_1994]. The benefit in using the top-down design is that the main focus is on the customers' requirement and the overall nature of the problem that should be solved.

In the bottom-up approach, the designers must identify a basic set of modules and their relations that can be used as the foundation for the problem solution [Thomas_1994]. Higher-level concepts are then formulated based on these basic elements. Bottom-up design is also an iterative process. *The benefit of the bottom-up design is that it permits the assessment and evaluation of the sub-modules during the system dsign process. But in the top-down design, performance evaluation can be done only when the complete system is integrated* [Thomas_1994]. However, top-down design does allow for early evaluation of functional capabilities at the user level by using dummy routines for lower-level modules. Thus, at the beginning of the project, the major interfaces can be tested, verified or exercised.

In practice, the pure top-down or bottom-up approach are seldom used. The top-down approach is best-suited when the problem and its environment are defined well. When the problem is ill-defined, the approach should mainly be bottom-up or mixed. The top-down approach have resulted in the evolution of a very popular design methods called structured design which will be discussed in the next section.

## 2.1.2 Structural Design Methods

Structured Design (SD) was first developed by Stevens, et. al [Stevens_1974]. It is easy to use and there is an evaluation criterion that can serve as a guide in the software design. The main notational scheme that SD uses is the data flow diagram (DFD). SD is based on tree important concepts: *composition* and *refinement* of the design; *separation* of issues into *abstraction* and *implementation*;

*evaluation* of the results. From the compositional point, SD views systems from two perspectives: as the flow of data and the transformations that data flow through a system. The designer can just focus on the transformations of the data flows through a system. Through the perception of the system as data flows and transforms, there is minimal variation in the construction of the system model, and as a result; the structure of the system is obtained. In addition, the interdependence of these data flows and transformations will result in the identification and organization of modules required in the building of the software system [EYourdon_1979].

From the abstract or implementation points, the SD process suggests a differentiation between the logical design and the physical design. Through the analysis of the data flows and the transformations (or sometimes called *mental analysis* [Budgen_2003]), the designer can derive a logical solution of the system. This early logical solution will not have details; will not be precise; and cannot be implemented immediately. Once this logical solution is able to satisfy the requirements or meet the objectives, the designer will then make the necessary changes so that the solution that can be implemented.

## 2.1.3 Object Oriented Design Methods

Object Oriented Design (OOD) methods provide a mechanism that has three important concepts in software design: *modularity*, *abstraction*, and *encapsulation* [Pressman_2001]. OOD is basically an approach that models the problem in terms of its *objects* and the operations performed on them. In OOD the system is decomposed into modules where each module in the system represents an object or class of objects from the problem space [Booch_1986]. Objects represent concrete entities which are instances of one or more classes. Objects encapsulate data attributes, which can be data structures or just attributes, and operations, which are procedures. A class is a set of objects that share a set of common structure and behavior. It contains three items: *class name*, *list of attributes*, and list of *operations*. The derivation of subclasses from a class is called *inheritance*. A subclass may have a few superclasses, thus multiple inheritance. The ability of any objects to respond to the same message and of each object to implement it appropriately is called *polymorphism*.

In Object Oriented Analysis (OOA) [Pressman_2001], a requirement analysis technique, starts at the top-level by identifying the objects and classes, their relationships to other classes, their major attributes and, their inheritance relationships then derive a class hierarchy from them. On the other hand, OOD extracts the objects, that are available from each class and their relationship to each other, to derive a detailed design representation. The basic building blocks to accomplish OOD are to establish a mechanism for: depicting the data structure; specifying the operation; and invoking the operation. OOD creates a model of the real world and maps it to the software environments. Even though OOD provides the mechanism to partition the data and its operations, its representations are prone to have programming language dependency. There is an absence of guidelines to model the initial objects or classes, thus it will depend upon analysis techniques from other methodologies. In the next section I am trying to focus on the *Rational Unified Process* as the Ideal frame work in software engineering.

## 2.2 Unified Framework for Software Design Methods

## 3. Software Quality

The quality of the design is about how well the design meets its requirements, or it could be defined as the "the *degree to which software possesses a desired combination of attributes*" [IEEE_Std_1061]. We should measure the attribute of the design to have better understanding of the model that we used. Different people may have different views about the quality of the design and how it could be measured. In some engineering discipline it is very easy to define the metrics that describe the performances of the system, for example when electrical engineers design on filter the quality of design could be defined about which range of frequency it supports, or how much power it dissipates. For software engineering defining the quality and metrics is a challenging job as it is mentioned by Fenton [Fenton_1991]:

> *"Measurement is the process by which numbers or symbols are assigned to the attributes of entities in the real world in such a way as to define them according to clearly defined rules. . . . In the physical sciences, medicine, economics, and more recently the social sciences, we are now able to measure attributes that we previously thought to be immeasurable. . . . Of course, such measurements are not as refined as many measurements in the physical sciences . . ., but they exist [and important decisions are made based on them]. We feel that the obligation to attempt to "measure the immeasurable" in order to improve our understanding of particular entities is as powerful in software engineering as in any discipline."*

Although defining the quality metrics and assessing the software quality is the challenging work in software engineering, in the next section I will try to present some frameworks that have been proposed for software design quality.

## 3.1 McCall's Quality Factors

The McCall's quality factors can be categorized into tree different aspects: operational characteristics, ability to undergo change and its adaptability to new environments. Fig. 4 shows these factors.

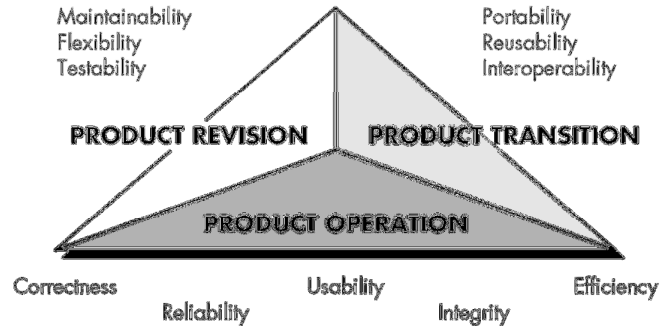

**Fig. 4 McCall's Software Quality Framework and Factors [Pressman_2001]**

They provide the following definitions for the factors [Pressman_2001]:

1. **Correctness**: The software satisfies its specification and customer mission objectives.
2. **Reliability**: The software can be expected to perform its intended function with required precision.
3. **Efficiency**: The amount of computing resources and codes required by a program to perform its function.
4. **Integrity**: The access to software or data by unauthorized persons can be controlled.
5. **Usability**: The amount of effort required to learn, operate prepare input and interpret output of the program.
6. **Maintainability**: Effort required locating and fixing an error in a program.
7. **Flexibility**: Effort required modifying an operational program.
8. **Testability**: Effort required testing a program to ensure that it performs its intended function.
9. **Portability**: Effort required transferring the program from hardware/software to the other environments.
10. **Reusability**: If the software component could be reused.
11. **Interoperability**: The effort required to couple one system to another.

In many cases it is very difficult to assign only one factor to one of these quality factors. If the linear relation is used we will have:

$$F_q = \sum_{i=1}^{N} C_i \times m_i$$

In which $F_q$ is software quality factor, $C_n$ are regression coefficient and $m_n$ is the metric that affects quality factor. The McCall represent the grading scheme from low (0) to high (10) scale. Fig 5. Represent the relation of the quality factor and some metrics. In this framework the check list will be used for assessing software quality.

| Quality factor \ Software quality metric | Correctness | Reliability | Efficiency | Integrity | Maintainability | Flexibility | Testability | Portability | Reusability | Interoperability | Usability |
|---|---|---|---|---|---|---|---|---|---|---|---|
| Auditability | | | | x | | | x | | | | |
| Accuracy | | x | | | | | | | | | |
| Communication commonality | | | | | | | | | | x | |
| Completeness | x | | | | | | | | | | |
| Complexity | | x | | | | x | x | | | | |
| Concision | | | x | | x | x | | | | | |
| Consistency | x | x | | | x | x | | | | | |
| Data commonality | | | | | | | | | | x | |
| Error tolerance | | x | | | | | | | | | |
| Execution efficiency | | | x | | | | | | | | |
| Expandability | | | | | | x | | | | | |
| Generality | | | | | | x | | x | x | x | |
| Hardware Indep. | | | | | | | | x | x | | |
| Instrumentation | | | | x | x | | x | | | | |
| Modularity | | x | | | x | x | x | x | x | x | |
| Operability | | | x | | | | | | | | x |
| Security | | | | x | | | | | | | |
| Self-documentation | | | | | x | x | x | x | x | | |
| Simplicity | | x | | | x | x | x | | | | |
| System Indep. | | | | | | | | x | x | | |
| Traceability | x | | | | | | | | | | |
| Training | | | | | | | | | | | x |

(Adapted from Arthur, L. A., *Measuring Programmer Productivity and Software Quality*, Wiley-Interscience, 1985.)

**Fig 5. Quality Factors and Metrics relation [Pressman_2001]**

For the definition of the metrics please refer to [Pressman_2001].

## 3.2 FURPS

This standard for software quality is defined by Hewlett-Packard [Grady_1987]. FURPS is the acronym of *functionality*, *usability*, *reliability*, *performance* and *supportability*. These factors are defined as [Pressman_2001]:

1. **Functionality:** The generality of the functionality of the program and its general security.
2. **Usability:** overall human factors, aesthetics, consistency and documentation.
3. **Reliability:** It is measured by the frequency of severity of failures, the accuracy of the outputs, ability to recover from errors, and program predictability.
4. **Performance:** It is measured by the processing speed, response time, throughput,….
5. **Supportability:** The ability to extend the program.

According to this framework of quality factors, metrics could be defined for measuring the quality of the software.

## 3.3 ISO 9126

This is the hierarchical software quality model that has been proposes for software quality. Fig. 6 describes this this model.

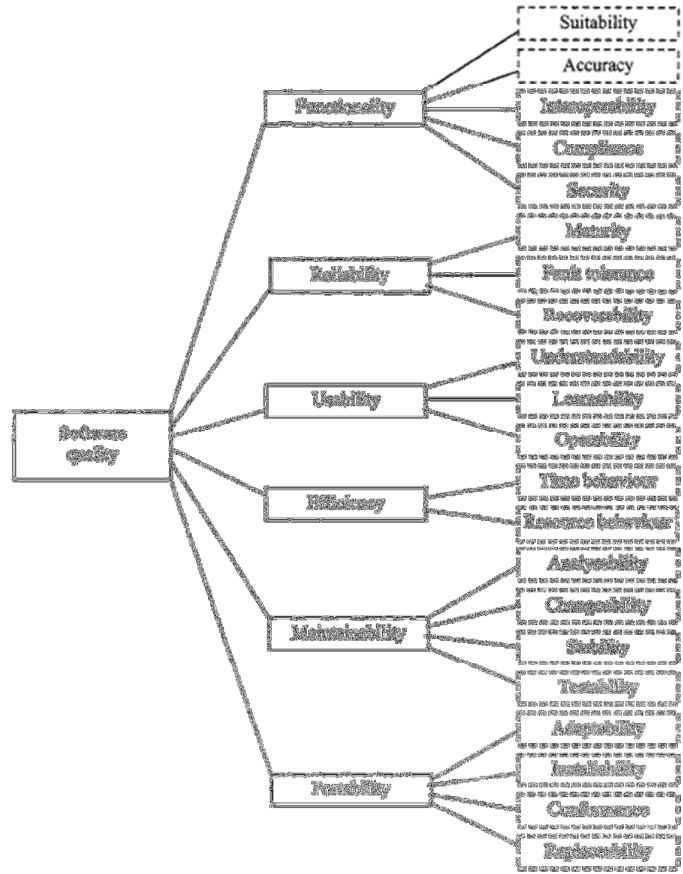

**Fig 6. ISO 9126 Software Quality Model [Zhu_2005]**

It can be understood from the above picture that the quality factors are dependent to some metrics that could be defined according to the specific software. For more information please refer to [Pressman_2001].

Now it is enough with the software quality factors, in the next section I will review some metrics that has been proposed in literature for measuring the software quality.